\documentclass[letterpaper,twocolumn,10pt]{article}
\usepackage{usenix,endnotes}
\usepackage{amssymb, amsmath, epsf, epsfig, graphicx, tikz}
\usepackage{bm}
\usepackage{xifthen}
\usepackage{color}
\usepackage{datetime}
\usepackage{framed}

\newenvironment{definition}[1][Definition]{\begin{trivlist}
\item[\hskip \labelsep {\bfseries #1}]}{\end{trivlist}}

\newcommand{\ee}[1]{\begin{align} #1 \end{align}} 						
\newcommand{\nn}[1][]{\ifthenelse{\isempty{#1}}{\nonumber \\}{\nonumber}}	

\definecolor{A}{cmyk}{0,0.48,0,0}
\definecolor{B}{cmyk}{0.79,0.88,0,0}
\definecolor{C}{cmyk}{0,0.52,1,0}
\tikzstyle{peers_A}=[circle, fill=black!15, draw, inner sep=0pt, minimum width=8pt]
\tikzstyle{peers_B}=[circle, fill=black!50, draw, inner sep=0pt, minimum width=8pt]
\tikzstyle{peers_C}=[circle, fill=black, draw, inner sep=0pt, minimum width=8pt]
\tikzstyle{peers_D}=[circle, fill=black!0, draw, inner sep=0pt, minimum width=8pt]
\tikzstyle{peers_AA}=[circle, fill=black!15, draw, inner sep=4pt, minimum width=8pt]
\tikzstyle{peers_BB}=[circle, fill=black!50, draw, inner sep=4pt, minimum width=8pt]
\tikzstyle{peers_CC}=[circle, fill=black, draw, inner sep=4pt, minimum width=8pt]
\tikzstyle{peers_DD}=[circle, fill=black!0, draw, inner sep=4pt, minimum width=8pt]
\tikzstyle{loop1}=                    [to path={
  \pgfextra{}
  [looseness=8,min distance=5mm,every loop]
  \tikz@to@curve@path
  }]
\tikzset{every loop/.style={}}

\begin{document}

\title{\Large \bf Symmetric Disclosure: a Fresh Look at $k$-Anonymity}
\date{}
\author{  {\rm EJ Infeld}
\\
Department of Mathematics, \\ 
Dartmouth College, Hanover, NH 03755, USA \\ 
ewa.j.infeld.gr@dartmouth.edu
 }

\maketitle

\subsection*{Abstract} 
We analyze how the sparsity of a typical aggregate social relation impacts the network overhead of online communication systems designed to provide $k$-anonymity. Once users are grouped in anonymity sets there will likely be few related pairs of users between any two particular sets, and so the sets need to be large in order to provide cover traffic between them. We can reduce the associated overhead by having both parties in a communication specify both the origin and the target sets of the communication. We propose to call this communication primitive \textquotedblleft symmetric disclosure.\textquotedblright$\ $If in order to retrieve messages a user specifies a group from which he expects to receive them, the negative impact of the sparsity is offset. 

\section{Introduction}

In the quest to build Internet communication systems that preserve the privacy of their users, many promising approaches have been proposed. Onion Routing remains the most popular and practical paradigm, and with good reason. Yet creating a solution that would compete with the most popular online services on efficiency while also providing provable privacy and security is still a challenge. Making users indistinguishable within certain groups, known as $k$-anonymity, would deliver strong protection against powerful adversaries \cite{KAN}, but it presents an array of practical hurdles. Many of those hurdles have been minimized in relevant research. 

Herbivore \cite{HERB} is a $k$-anonymity based system that can serve as a point of reference in this discussion. By dividing users into smaller groups, Herbivore manages to reduce resource overhead significantly. We will build on and expand this discussion. Both Herbivore and, more recently, Dissent \cite{BUDDIES, DISSENT, PANOPTICON} use DC-nets to make sure that their anonymity sets are reliable. They mark admirable progress in making sure the architectures scale. Aqua \cite{AQUA} is another recent model, which groups clients that exhibit similar behavior to achieve $k$-anonymity. Aqua maintains these anonymity sets by using mixes, and has a weaker threat model than Herbivore and Dissent, but limits the bandwidth overhead. The strategy we propose in this paper could be implemented straightforwardly in a scheme such as Aqua.

We think of a social network or large communication system as a graph where nodes correspond to the users, and edges to a social relationship such as Facebook's \textquotedblleft friendship.\textquotedblright$\ $ We model a communication between users as traveling over an edge, from a user to one of his friends and call all communications \textquotedblleft messages.\textquotedblright$\ $ We then group users into $m$ anonymity sets, by assigning each user to a set uniformly at random. These sets will have close to $n/m$ users, and will be extremely unlikely to have significantly less users, so we can pick $k=n/2m$. This grouping conceals the social graph from the service provider, and if methods of existing powerful $k$-anonymity schemes \cite{HERB,DISSENT,AQUA} are employed, from powerful global adversaries.

This paper sets out to discuss group sizes and resource trade-offs for $k$-anonymity systems. As a point of departure, we look at the amount of traffic between two randomly chosen groups. Since a typical social relation is sparse, this amount is small relative to the size of the groups. For any connection between two groups, any other connection between the same groups can serve to provide cover traffic to hide communication patterns on that particular connection. As the amount of expected connections grows, so do the sizes of the groups. In fact in the case of Facebook there would need to be on the order of thousands of users in each group in order for us to have a reasonable expectation of having any cover traffic at all. In a setting where messages meant for each member are broadcast to the entire group, this would mean that we need to accept high bandwidth overhead in order to gain cover traffic. We argue that if each member of a group only retrieves communications from a group in which he has a friend, no information about the social graph is compromised.

Different services may require different amounts of cover traffic, and therefore different group sizes. We divide our discussion into consequences of opting for little resource premiums (\textit{light} design), and for a high probability of achieving a specific amount of cover traffic (\textit{stream} design). We then borrow some intuition from differential privacy to argue that perhaps a middle of the road design that we call \textit{hybrid} strikes a good balance. The aim of the \textit{hybrid} design is to make the groups just big enough to be likely to provide \textit{some} cover traffic.

We open with an intuitive introduction in Section 2. We start by describing the general idea of a $k$-anonymous system with multicast, and then argue for discriminating between messages based on their source group. Section 3 provides some relevant formal definitions. In Section 4, we estimate desired sizes of the groups based on the social graph of Facebook. Section 5 looks at real-world practicality and resource use, and Section 6 suggests a direction that could further reduce the resource premiums. 
We present our conclusions in Section 7.

\section{Key idea}

Let us start by painting an intuitive picture of what we are trying to do. For the purposes of a theoretical model, we envision a centralized communication service. The same structure, however, could be implemented in a decentralized way. Our goal is to allow the users to communicate while concealing the structure of the social graph.

First, a user signs up for the service and gets assigned to a group uniformly at random. Unlike in Aqua \cite{AQUA}, no individual properties of the users are a factor in the group assignment. It's worth noting that Herbivore \cite{HERB} uses an assignment protocol that attempts to keep groups approximately evenly sized while also taking the decision out of the users' hands. We have the same goals, but choose to simplify the protocol for the sake of this theoretical model by keeping the number of groups fixed and assigning each user to a group uniformly at random (and in particular \textit{not} according to local properties or relationships on the social graph). At this point the user's identity is not hidden.

 Suppose that from then on, a user is reliably hidden within his group. In practice, this might mean that the users interact with the service on their own as a part of a DC-net or through a system of mixes. The latter could be either Aqua's architecture that assumes entry and exit mixes are not compromised, or we could suppose the users connect to the system through Tor and we choose to disregard potential de-anonymizing attacks for the purposes of this paper. In both cases, the users maintain $k-1$ anonymity with respect to other members of their group. The computational requirements of authenticating in a way that would keep a user anonymous within their group are tractable, as there has been remarkable progress in the field of anonymous authentication techniques such as linkable ring signatures \cite{LINK1, LINK2}. They require constant time computation and allow the system to revoke misbehaving users \cite{PERM}. 
 
 In order to send a message, a user will first encrypt it with the intended recipient's individual public key, and then enter it to the system as a message addressed to the recipient's group. So if Alice is in group $A$ and Bob is in group $B$, and Alice wishes to send a message to Bob, she encrypts it with Bob's public key, and tells the service \textquotedblleft I am a member of group $A$ and this is a message for group $B$.\textquotedblright$\ $ The source and destination group of the message will be known, so we can view the message as equipped with a pair $(A,B)$ as a message header. So far, we're describing a familiar idea of a $k$-anonymous system with multicast.
 
 For Alice to find out Bob's group and key, either a separate system could be in place for establishing the friendship relation as in Pond  \cite{POND}, or we could use a public address book such as Cryptobook \cite{BOOK}.

A typical social relation is sparse. A social network can have hundreds of millions of users and each will only have a few hundred friends. Unless the groups we put users in are big, if Alice and Bob are friends they are likely to be the only connection between their groups. Suppose that Alice sends a message to Bob. What she reveals is that there is a connection between her group and Bob's. Bob knows he has a friend in Alice's group. If, instead of retrieving all messages that is meant for his group he only retrieves \textquotedblleft messages for group $B$ that originated in group $A$,\textquotedblright$\ $ he's not revealing extra information about the social graph - the connection between groups $A$ and $B$ has already been revealed by Alice. And thanks to the sparsity, he's unlikely to have to download any messages that are not meant for him, as if Bob's group had a separate post box for each sender group. 

\begin{figure}[ht!]
\centering
\includegraphics[width=63mm]{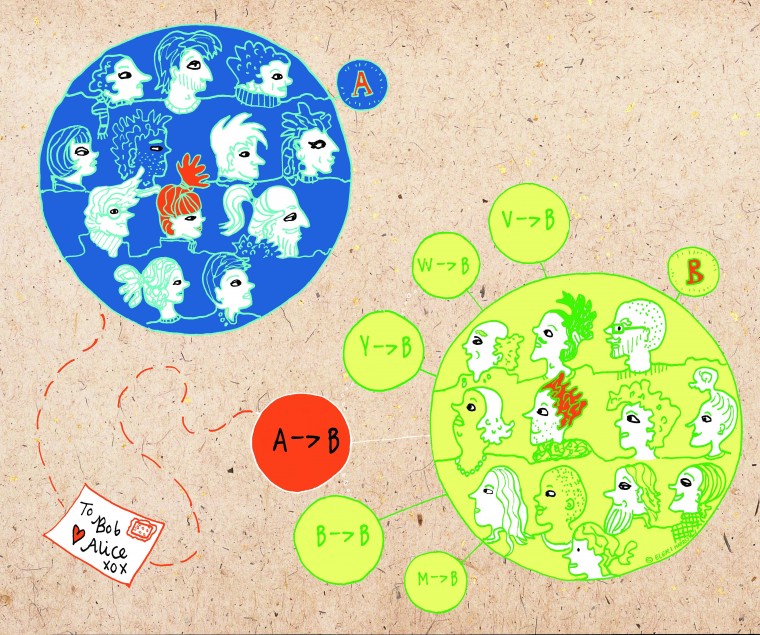}
\caption{Bob only opens the post box with messages that arrived from Alice's group.\hfill \tiny{Illustration \textcircled{c} Eleri Mai Harris 
2014}}
\label{fig:postboxes}
\end{figure}

This change could help cut down on broadband in a system that employs mixes, such as Aqua \cite{AQUA}. Aqua was designed to anonymize BitTorrent traffic, its primary role is to facilitate download. In this type of an architecture, Bob can specifically request the messages meant for group $B$, originating in group $A$. As long as Bob's requests for messages from separate groups can't be linked, he doesn't need to employ a PIR scheme and the social graph will remain obscured. We come back to the question of unlinkability in Section 5.

Discriminating between messages based on source group may not cut down on bandwidth in systems that employ DC-nets, since they require all communications to be broadcast to the entire group as part of their functioning. In such case the most obvious strategy of broadcasting every message to all member exactly once remains optimal. However, our discussion is valid for the purposes of computation cost - Bob can discard messages that come from groups he has no friends in without trying to decrypt them. In this case, the estimates in Section 4 remain valid.

This is what we refer to as \textit{symmetric disclosure}: the fact that a message \textit{exists} already reveals the fact that there is a connection between groups $A$ and $B$. Surely, the person it is meant for is one of the members of group $B$ that do in fact have a friend in group $A$, and so Bob disclosed information that had already been revealed by Alice. A standard multicast method would have Bob dealing with every message addressed for his group. If there are $m$ groups, Bob would need to sieve through on average one in $m$ messages sent in the entire network. But once we shift focus to pairs of groups, such as $(A,B)$, this is on the order of $1/m^2$ for each of Bob's friends. The trade-off is that if we use this fact to save on broadband, the adversary learns a more narrow set of possible pickup times for the message. 

For $k$-anonymity, the sparsity of a social relation seems like a problem - even in a $k$-anonymous setting, it might be easy to distinguish the traffic between a specific pair of users that are known to be friends. Symmetric disclosure is an attempt to neutralize the cost of this sparsity. If used to its full potential, we need only give up as much bandwidth as much cover traffic we are trying to gain, and can set the group sizes accordingly.

\subsection{Threat model}

On its own, the presented scheme is designed to do well against a passive observer with the capabilities of a service provider, or as an anonymized dataset. With every encrypted communication between two users, the adversary learns the source group, send time, destination group, and a set of potential pickup times as well as the size of the communication.

However, the same discussion applies to powerful $k$-anonymity systems \cite{HERB, DISSENT, AQUA} that would allow us to assume the users are reliably hidden within their groups, and in particular in Aqua \cite{AQUA} \textit{symmetric disclosure} could be easily implemented. In such a case we will assume the threat model of Aqua: an adversary that is also capable of global traffic analysis, taking over parts of the system's infrastructure, and modifying communications. The Aqua threat model also allows for a bounded number of compromised clients. We could rely on the revocation capabilities of the login protocols to prevent such attacks, but we do not discuss it in this paper. Any adversary is computationally bounded in a way that allows for use of public key encryption.

\section{Theoretical setup}

Now that we have an intuitive idea of what we are trying to achieve, let's put it on a formal footing. We define a communication network as a set of users equipped with a friendship relation, and content traveling from one user to another. For simplicity, call all content \textit{messages} and assume that these messages are evenly sized.

\begin{definition}Let a communication network $G=(V,E,M)$ consist of the set of $n$ users $V$, a friendship relation $E$ between the users and the set of messages $M$. If $a,\ b\in V$, a message $m\in M$, from $a$ to $b$ is defined by quintuple $(a,\ b,\ t,\ t',\ c_{b})$ where $t$ is the time the message is sent, $t'$ is the time the message is received and $c_{b}$ is the message's content encrypted for $b$.
\end{definition}

We model the underlying social graph $(V,E)$ the usual way, with the nodes in the graph representing the users, and Alice and Bob connected by an edge if and only if they are friends in this network. We will partition the set of nodes $V$ it into $m$ parts, where $m$ is a parameter to be determined based on anonymity preferences. We do this by allocating each of the $n$ users to one of the groups $\{V_1,V_2,\dots,V_m\}$ uniformly at random. These groups will serve as stable anonymity sets for the users. 

We arrive at a projection operator $\mathcal{P}_m:G\rightarrow \mathcal{G},$ that sorts the users into equivalence classes $\{V_1,V_2,\dots,V_m\}$, and projects the friendship relation $E$ to a relation $\mathcal{E}$ between the sets $\{V_1,V_2,\dots,V_m\}$ defined by $\{V_i,V_j\}\in \mathcal{E}$ if and only if $\{x,y\}\in E$ for some $x\in V_i$ and $y\in V_j.$

Suppose that there is a message traveling from $a\in V_A$ to $b\in V_B$. From the point of view of the service, a user declared themselves to be \textit{a member of group $V_A$ wishing to send a message to group $V_B$}, then one or more users declared themselves to each be \textit{a member of group $V_B$ wishing to download messages incoming from group $V_A$}. 

Under the action of $\mathcal{P}_m$, $\ \mu=(a,\ b,\ t,\ t',\ c_b)\in M$ becomes $\mathcal{P}_m(\mu)=(V_A,\ V_B,\ t,\ \mathcal{T}_{V_j}',\ c_{V_B})\in\mathcal{M}$ where $V_A$ and $V_B$ contain $a$ and $b$ respectively, and where $\mathcal{T}_{V_B}'=\{t_1',t_2',\dots,t_{|V_B|}'\}$ is the set of (unassigned) times at which the message $\mathcal{P}_m(\mu)$ is received by members of $V_B$. Most of the entries will be empty, as most members of $V_B$ will never receive the message.

\begin{definition}Let a projection network $\mathcal{G}=(\mathcal{V,E,M})$ consist of the set $\mathcal{V}$ of $m$ user equivalence classes, $\{V_2,V_2,\dots,V_m\}$, a relation $\mathcal{E}$ between the sets and the set of messages $\mathcal{M}$.
\end{definition}

\begin{figure}[ht!]
\hfill
\begin{tikzpicture}[scale=0.3]
\foreach \place/\name in {{(-2,-1)/a}, {(-4,3)/c}, {(-5,5)/k}, {(4.5,-3)/o}}
    \node[peers_A] (\name) at \place {};
\foreach \place/\name in {{(2,1)/b}, {(2,3)/e}, {(3,5)/f}, {(2,-3.5)/n}}
        \node[peers_B] (\name) at \place {};
  \foreach \place/\name in {{(-1,4)/d}, {(4,-1)/h}, {(-2,5)/l}, {(0,-2)/p}}
    \node[peers_C] (\name) at \place {};
  \foreach \place/\name in {{(5,4)/g}, {(3,-2)/i}, {(-6,1)/j}, {(-7,-3)/m}}
    \node[peers_D] (\name) at \place {};
\foreach \source/\sink in {a/c, d/a, c/b, d/c, d/e, b/i, b/h, b/e, e/g, e/f, g/f, d/e, j/c, k/c, l/c, j/a, j/m, n/o, n/i, o/i, p/b}
    \path (\source) edge (\sink);
\draw[text=black] (8,1.5) node {$\mathcal{P}_4$};
\draw[text=black] (8,0) node {$\longrightarrow$};
\foreach \place/\name in {{(13,2)/a}}
    \node[peers_AA] (\name) at \place {};
\foreach \place/\name in {{(17,2)/b}}
        \node[peers_BB] (\name) at \place {};
  \foreach \place/\name in {{(16,-1)/d}}
    \node[peers_CC] (\name) at \place {};
  \foreach \place/\name in {{(12,-2)/g}}
    \node[peers_DD] (\name) at \place {};
      \foreach \source/\sink in {d/a, a/b, d/b, b/b, b/g, a/g, a/a, g/g}
        \path[thick] (\source) edge (\sink);
\path[thick] (a) edge [loop] node {} (a);
\path[thick] (b) edge [loop] node {} (b);
\path[thick] (g) edge [in=225, out=315, loop] node {} (g);

\end{tikzpicture}\hfill\hfill
\caption{The action of projection operator on a graph.}
\label{fig:awesome_image}
\end{figure}
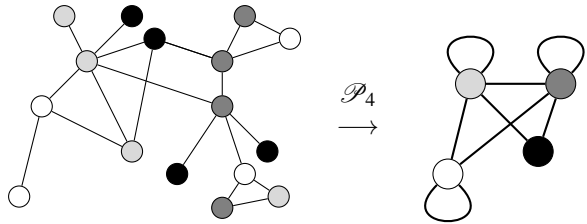

One might argue that a standard multicast model allows for a bigger set of possible pickup times of a particular message. But as in all cases where the receiver is hidden by a stronger mechanism than a sender, such advantage is voided in back-and-forth correspondence.

\section{Group sizes}

If Alice and Bob are the only connection between their groups, they won't use any more bandwidth or computational power to communicate than if they were using an encrypted communication network without a projection. But this still allows for pattern analysis - say, if an adversary has auxiliary information that suggests that Alice and Bob's communications are of interest, he can find valuable information.

In some settings this could be considered a feature. If two specific people are of interest to a law enforcement agency and known to be likely to communicate, with small enough groups their apparent traffic can be analyzed, however circumstantial. The good news is that both parties need to be of interest in the first place - simply communicating with someone who is of interest will not put a user under suspicion. Likewise, allowing for analysis of such anonymized patterns could work well in certain research settings.

We can therefore choose to set the number of groups to be large enough that Alice and Bob are likely to be the only pair of friends between their respective groups in the projection network, or alternatively, small enough so that there are likely to be other people communicating between these two groups. The first design is efficient, while the second has a chance of protecting the users from traffic analysis - the choice would depend on the intended nature of the service. We divide our discussion into the following two settings:\medskip

\textbf{\textit{Light} design:} There are sufficiently many groups so that a connection between groups is likely to be unique.\medskip

If an adversary has no information other than the data in the network that suggests that Alice might be the one sending messages to Bob's group, she will be hidden in a powerful anonymity set at little cost. In section 4.1 we find that if we applied this division to Facebook's social graph the anonymity sets could be of average size about 916 users and only 20\% of messages would be downloaded more than once.\medskip

\textbf{\textit{Stream} design:} In this case, we adjust the number of groups so that there are few pairs of groups that have less that $l$ connections between them. \medskip

\textbf{\textit{Hybrid} design} This is the \textit{stream} design with $l=1.$\medskip

We can approximate the number of edges between groups by a Poisson distribution. Suppose we decide that a particular minimum $l$ should be attained with probability $1-\epsilon$. If the network has millions of users, we expect this to result in several thousand of users in each group. We are effectively constructing anonymity sets for connections. Let us call the combined connections between two groups $V_A$ and $V_B$ the \textit{stream} between $V_A$ and $V_B$. Depending on the context, we can choose how much efficiency we are willing to give up to ensure that an edge has a stronger anonymity set.

\subsection{Estimating group sizes based on Facebook's social graph}

The schemes we just identified point to a desired probability distribution of the number of connections between a pair of groups chosen uniformly at random. We now set out to estimate the sizes of anonymity sets that these might yield in a large social graph. For our evaluation we will approximate the degree distribution of the Facebook social graph as described by Ugander et al. \cite{ANA}, and the accompanying dataset \cite{FACE}. It is worth noting that since users are assigned to groups uniformly at random, we do not need to consider clustering properties of the friendship relation, only the distribution of degrees of particular nodes. The mean degree is $d=191.4161$, with standard deviation $\sigma=190.4263$, and $n=721094633$.

In order to gauge the group sizes in \textit{light} and \textit{hybrid} designs, we would like to estimate the probability that Alice and Bob, while corresponding, will have to deal with messages other than their own. In the \textit{light} design, we would like to minimize this probability. In the \textit{hybrid} design, we would like to maximize it while still keeping the mean number of connections between two groups low for the sake of efficiency. 

The probability that Alice and Bob, while corresponding, will have to deal with messages other than their own is the probability that while the two of them are connected with an edge in $G$, there is \textit{another} edge between $V_A$ and $V_B.$ Due to the sparsity of the friendship relation in the graph, the probability of any particular edge existing is negligible (on the order of $d/n$). So after we remove the edge \{Alice, Bob\}, the probability that $V_A$ and $V_B$ are connected is equivalent to the probability that, if a pair of groups is picked uniformly at random, there exists an edge between these groups.

For any given groups $V_i$ and $V_j$, we estimate the probability that there exist no edges between them. Even though the degree structure is preserved in the model, on the relevant values of $m$ this can be well approximated by a Poisson random variable with parameter $\frac{nd}{m^2}$ that corresponds to distributing edges between pairs of groups. (The total number of edges is $nd/2$, the total number of pairs with repetition (we allow edges with both ends in the same group) is $m^2/2$.) We will present a more thorough evaluation below, but we find that the two distributions agree numerically for all typical values of $m$.

We can describe the number of users in a group with a binomial distribution with mean $n/m.$ This can be well approximated by a normal random variable with mean $n/m$ and standard deviation $\sqrt{n/m}$. 

Suppose that we are looking at a group $V_i$, with $|V_i|$ nodes. We can use the Central Limit Theorem to approximate the distribution of the sum of node degrees in a particular group $V_i$ as normal distribution with mean $d\times |V_i|$ and variance $\sigma^2\times|V_i|$.

\begin{figure}[ht!]
\centering
\includegraphics[width=85mm]{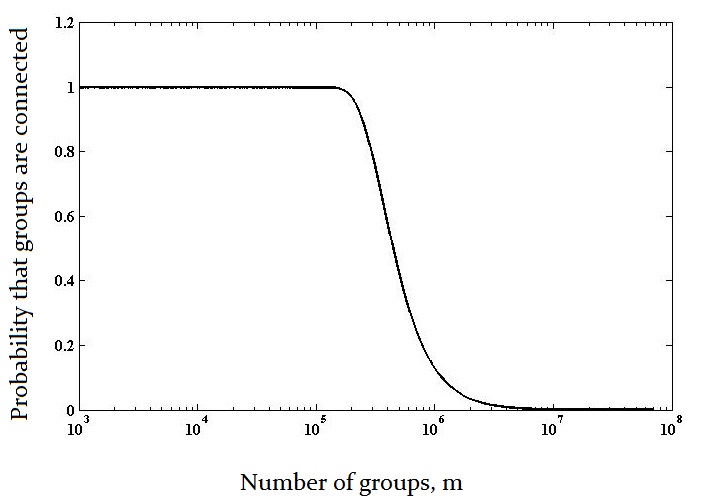}
\caption{The probability $1-\epsilon$ that two groups are connected.}
\label{fig: 1con}
\end{figure}

We can conclude that for a group $V_i$ the probability that it has $j$ members and the sum of their degrees is $D$ can be described by $\mathcal{N}(j;\frac{n}{m},\sqrt{\frac{n}{m}})\times \mathcal{N}(D;jd,\sqrt{j}\sigma).$ Every edge with one end in $V_i$ has chance $\frac{1}{m}$ of its other end being in $V_j$, so the event that none of these land in $V_j$ hes probability $(1-\frac{1}{m})^D$. Then the probability $\epsilon$ that for randomly picked $V_i,\ V_j$ we have $\{V_i,V_j\}\notin \mathcal{E}$ is:\scriptsize
\ee{\mathbb{P}[\{V_i,V_j\}\notin   \mathcal{E}]\simeq \underset{\mathbb{R}^+\times\mathbb{R}^+}{\int}\!\!\mathrm{d}x&\ \mathrm{d}j \ \left(1-\frac{1}{m}\right)^{x}\times\mathcal{N}(j;\frac{n}{m},\sqrt{\frac{n}{m}})\times \mathcal{N}(x;jd,\sqrt{j}\sigma)\ .}\normalsize 
The graph in Figure~\ref{fig: 1con} plots the probability $1-\epsilon$ that a pir of groups picked uniformly at random is connected, with respect to parameter $m$. The numerical values for the corresponding Poisson distribution in the drop off region are in excellent agreement with ones obtained in this way. We find from Table$\ $ ~\ref{tab: probabilities} that for group sizes of $n/m\sim916$ only about 20\% of connections are not unique, and for $n/m\sim194$ this becomes about 1\%. The values at the top of Table \ref{tab: probabilities} correspond to the \textit{hybrid} design, while those at the bottom correspond to the \textit{light} design.

Thanks to the properties of the Poisson distribution, we can evaluate group sizes in the \textit{stream} design in much the same fashion, and the probability that two groups are connected by at least $l<< m$ edges will yield the same variance.

\begin{table}[ht!]
\caption{Value of $m$ and group size $n/m$ for various $1-\epsilon$}
\centering
\begin{tabular}{| c | c | c | c |}
\hline\hline\\[-2ex]
\ \ \ $1-\epsilon$\ \ \  &\ \ \ \ \ \ \ \ \ $m$\ \ \ \ \ \ \ \ \  &\ \ \ mean \ \ \ & \ \ \ mean \# \ \ \ \\ & & group size & connections \\ [0.5ex] 
\hline\\[-2ex]
0.999&$\sim$141,360&$\sim$5100 & 6.9077 \\
0.995&$\sim$161408 & $\sim$4467 & 5.2963 \\
0.99&$\sim$173,125&$\sim$4165 & 4.6052 \\
0.9 & $\sim$244,837 &$\sim$2945 & 2.3026 \\
0.8 & $\sim$292,852 & $\sim$2462 & 1.6094 \\
0.75 & $\sim$315,542 & $\sim$2285 & 1.4863 \\
0.2 & $\sim$786,490 &$\sim$916 & 0.2231 \\ 
0.1 & $\sim$1,144,580 &$\sim$630 & 0.1054\\
0.01 & $\sim$3,705,910 &$\sim$194 & 0.0101 \\
\hline
\end{tabular}
\label{tab: probabilities}
\end{table}

\subsection{Edge privacy of $\mathcal{E}$}

Let us provide an additional bit of intuition than might guide a choice of target amount of traffic. Suppose that Alice sends a message to Bob, but the adversary doesn't know that Alice is a person of interest. What she discloses to the network is that there is a connection between group $V_A$ - her group, and group $V_B$ - Bob's group i.e. that $\{V_A,V_B\}\in \mathcal{E}$. We would like this information to be statistically meaningless. 

For example, suppose that Alice an Bob are friends and that $1-\epsilon=0.99.$ An adversary obtains the graph of $\mathcal{G}=(\mathcal{V,E})$, i.e. a projection network of the system. The probability that if we remove Alice from the dataset it will still be the case that $\{V_A,V_B\}\in \mathcal{E}$ is 0.99.

We will use the following definition of edge-privacy:
\begin{definition} 
Suppose that Alice is in group $V_A$ and Bob is in group $V_B$ and $\{\text{Alice, Bob}\}\in E$. The model is $\epsilon$-edge-private for Alice sending a message to Bob if there exist an edge between $V_A-\{\text{Alice}\}$ and $V_B$ with probability at least $1-\epsilon.$
\end{definition}

This definition is not symmetric. In the spirit of differential privacy, we are examining the chances of getting $\{V_A,V_B\}\in \mathcal{E}$ in the projection network $\mathcal{G}$ after changing the information of a single agent, Alice. $1-\epsilon$ is the probability that Alice and Bob are downloading extra messages when retrieving messages from each other. The \textit{hybrid} design was the one where we tried to maximize this probability while keeping the mean number of connections low. We might therefore argue that if building a $k$-anonymity based system one might consider aiming at groups that are just big enough to preserve edge-privacy.

\subsection{Traffic analysis and group reassignment}

Hopper and Vasserman \cite{TRAFFIC} analyzed the resistance of $k$-anonymity to global traffic analysis and concluded that if the groups change rarely, statistical disclosure attacks will fail. They point out that this is problematic in practice - in a practical system, groups are likely to fluctuate according to users' behavior. They quantify the acceptable \textit{churn} of the network. Applying their formula to what we called a hybrid design for the Facebook statistics, we can accommodate on the order of $10^4$ group changes, with an additional factor depending on the desired confidence parameter.

We have not considered the growth of the network, but instead looked at a stable social graph with $n$ nodes. Locally, the we can handle the growth of the graph by simply assigning a new user to one of $m$ pre-existing groups as she joins the network. 

In the long run the number of sets $m$ needs to be updated as the network grows, if we are hoping to keep the target density in the projection network. At this point the users need to be assigned to groups again uniformly at random, or groups could be split into smaller ones as it is done in Herbivore \cite{HERB}. Small networks are more dense since the user base tends to be less diverse. The number of sets we need will grow slower than the network.

\section{Bob's query quandary}

It is important that Bob's requests for messages from separate groups can not be correlated and linked. This correlation would allow an adversary to fingerprint Bob, and over time collect such lists for most users, thus recovering much of the graph structure. 

In a decentralized system using mixes, Bob would most likely receive his mail from each group over a separate mix path. We must take care that that these requests are not correlated based on timing. In a centralized setting, this danger is even higher. Suppose then that Bob disguises his IP address separately for the purposes of each group he needs to request messages from. Suppose that Bob has friends in $d_B$ groups. A conservative strategy of connecting at random times and checking a random friend group each time would result in $\Theta(d_B\log(d_B))$ checks as in the coupon collector's problem.

If Bob prefers connecting only once, he could check all messages meant for his group. He can then discard whatever he knows is not of interest. Even if the architecture of the system allows for symmetric disclosure, it may not be a preferable strategy in small, comparably dense networks. We will call the one connection scenario the \textit{bulk download} (BD) option for each scheme, and include if for comparison.

There are three types of resource use that Bob is concerned with:\small \begin{itemize}\item Bandwidth use, measured in the ratio of the messages Bob is likely to download over ones actually meant for him. \item Computational power, in decryption attempts Bob has to perform over the number of messages meant for him. \item Time, measured in the number of connections he needs to make to the service to check all of his messages.\end{itemize} \normalsize In the light design, as long as Bob checks his messages from one friend group at a time, there are negligible premiums on bandwidth and computation. Similarly, in the hybrid design these premiums are small enough factors that they could be handled by a personal computer without impact.

Table 2 compares the resource use in these scenarios. For the sake of computational efficiency and as an anti-spam measure, assume that only a small part of a message needs to be downloaded and analyzed for Bob to determine whether the message is meant for him. This could be done with a small encrypted header, that would then serve as an implicit address and include a designated-verifier digital signature \cite{VERIFIER} so that Bob can be sure that the message is not only meant for him, but comes from Alice. In the table, we assume that the headem measures $\varphi$ of a message size. We don't consider PIR protocols, but they could be useful in medium-sized networks.

Let $d_B$ be the number of groups Bob has friends in, $\varphi$ the ratio of the size of the identifying appendix to message size, and $\mu_\epsilon$ mean number of connections between two groups in hybrid and stream design. In hybrid, a possible value of $\mu_\epsilon$ might be about 5 or 7 for an $\epsilon$ of 0.005 or 0.001, as evidenced in Table 1. Unless the distribution of node degrees in the graph is highly anomalous, this number depends only on the variance of Poisson distribution, rather than size or density of the graph itself.

\begin{table}[ht!]
\caption{Approximate resource use}
\footnotesize
\centering
\begin{tabular}{|c | c | c | c |}
\hline\hline\\[-2ex]
Scheme &  Light & Stream/ &  BD\\ 
& & Hybrid & \\
\hline
Bandwidth& $1$ & $\mu_\epsilon(1+\varphi)$ & $|V_B|(1+\varphi)$ \\[0.5ex] 
Computation& $1$ & $1+\varphi\mu_\epsilon$  & as non-BD \\[0.5ex] 
\# connections  & $\Theta(d_B\log(d_B))$  & $\Theta(d_B\log(d_B))$  & $1$  \\[0.5ex] 
\hline
\end{tabular}
\end{table}

\section{A possible improvement}

The number of connections that Bob has to make to check all of his messages could prove to be a time issue. Let us explore what one can do to minimize it.

So let us again look at a situation in which Alice is in group $V_A$ and Bob is in group $V_B.$
If another user, Charlie, retrieves messages that go from Alice's to Bob's group, no harm is done to Alice or Bob.
All the addressing that is specific to Bob is implicit and so won't be visible to Charlie, he might even provide a useful decoy. 

As Alice writes to Bob, she could append as part of the content of her message that the next message will be sent to $V_Y$. Bob will retrieve it as sent from $V_A$ to $V_Y$. Later, Alice may change her strategy. This alone can serve to conceal the patterns in Bob's retrieval requests. 
If the sender groups could also be made dynamic in this way, we could all but eliminate the resource premiums. It could prove to be an interesting quandary to explore what kind of a protocol minimizes the number of sender groups Bob needs to retrieve messages from.

\section{Conclusions}

While some efficiency concerns remain, in particular with respect to the number of separate queries each user has to make, we believe that this approach has merits. Building a provably anonymous communication system is an important problem and $k$-anonymity might be the most promising take on it so far. Symmetric disclosure is step in bringing it to the realm of feasibility. This idea promises, in its \textit{light} form, a practical proposal to limit useful information obtained from large communication datasets to communication patterns of pairs of people that are known to communicate due to auxiliary information. It also provides a good way to obtain research datasets of communications without exposing the parties in any particular communication. In its \textit{hybrid} and \textit{stream} form, symmetric disclosure hides the communication patterns in anonymity sets of their own.

In both cases, it sets out to limit the network overhead to the amount of cover traffic that would be obtained and no more.

\section{Acknowledgments}
The author would like to thank Bryan Ford for terrific guidance as shepherd, and the FOCI reviewers for valuable input. Many thanks to Gwen Spencer, Zachary Hamaker, Damian Sowinski, Peter Johnson, Fran\c{c}ois Dorais, 
Pete Winkler, and to the very talented illustrator Eleri Harris (elerimai.com) for her rendition of symmetric disclosure in Fig.$\ $ ~\ref{fig:postboxes}.

\end{document}